\documentclass[conference,letterpaper]{IEEEtran}
\IEEEoverridecommandlockouts
\usepackage{cite}

\usepackage{array}
\usepackage{amsmath}
\usepackage{amssymb}
\usepackage{graphicx}
\usepackage{multirow}
\usepackage{stfloats}
\usepackage{textcomp}
\usepackage{url}

\begin{document}

\title{Hypergraph-Enhanced Dual Convolutional Network for Bundle Recommendation}

\author{
\IEEEauthorblockN{Yang Li$^{1\ast}$, \IEEEmembership{Member, IEEE}, Kangbo Liu$^1$, Yaoxin Wu$^2$,\\
Zhaoxuan Wang$^1$, and Erik Cambria$^3$, \IEEEmembership{Fellow, IEEE}}
\IEEEauthorblockA{$^1$Northwestern Polytechnical University, Xi'an, China\\
$^2$Eindhoven University of Technology, Netherlands\\
$^3$Nanyang Technological University, Singapore\\
$^\ast$Corresponding author: liyangnpu@nwpu.edu.cn}
}

\maketitle

\begin{abstract}
Bundle recommendation ranks sets of related items rather than isolated items. Its central challenge is to connect user preferences, item interactions, and bundle composition without losing the signals needed to rank bundles. We propose Hypergraph-Enhanced Dual Convolutional Network (HED), which constructs a complete hypergraph containing user--bundle, user--item, and bundle--item interactions together with intra-user and intra-bundle relations. HED couples complete-hypergraph propagation with a user--bundle branch, allowing item-aware higher-order context to inform ranking while preserving recommendation-specific signals.
On NetEase, HED-128 improves over the strongest baseline by 5.04--6.97\% across the six reported metrics; on Youshu, HED-64 improves by 1.87--4.56\%. Ablation results support the contributions of both the user--bundle branch and intra-type relations, and sensitivity analyses identify stable operating ranges for the main hyperparameters. We further quantify the computational trade-off of the complete hypergraph, including its memory cost. The evidence supports HED on the two evaluated bundle-recommendation datasets while making its resource limitations explicit.

\end{abstract}


\section{Introduction}
Recommender systems identify items that are likely to interest users from large collections such as books, music, videos, and games~\cite{li2016,li2017Price,li2019,chen2023bias,zheng2023automl}. Their effectiveness depends on learning user preferences from item characteristics and historical interactions such as purchases or clicks~\cite{10494051}, a setting closely tied to the broader analysis of large-scale social and behavioral data~\cite{cambig}. Personalization driven by user preferences and affect also underpins emotion-sensitive and patient-centered systems~\cite{camsen,campat,campro}.

Many online platforms also recommend bundles: sets of items with internal membership or affiliation relationships. Examples include digital game stores and online music services, where bundles can simplify selection for users and support package-oriented sales. The prevalence of such bundles motivates bundle recommendation as a distinct task. In particular, Deng et al.~report that discounted bundles account for a substantial share of revenue in a mobile game~\cite{deng2020personalized}. The central modeling challenge is to exploit the interactions among users, items, and bundles without discarding the structure that links them.

\begin{figure}[h]
	\centering
	\includegraphics[width=\linewidth]{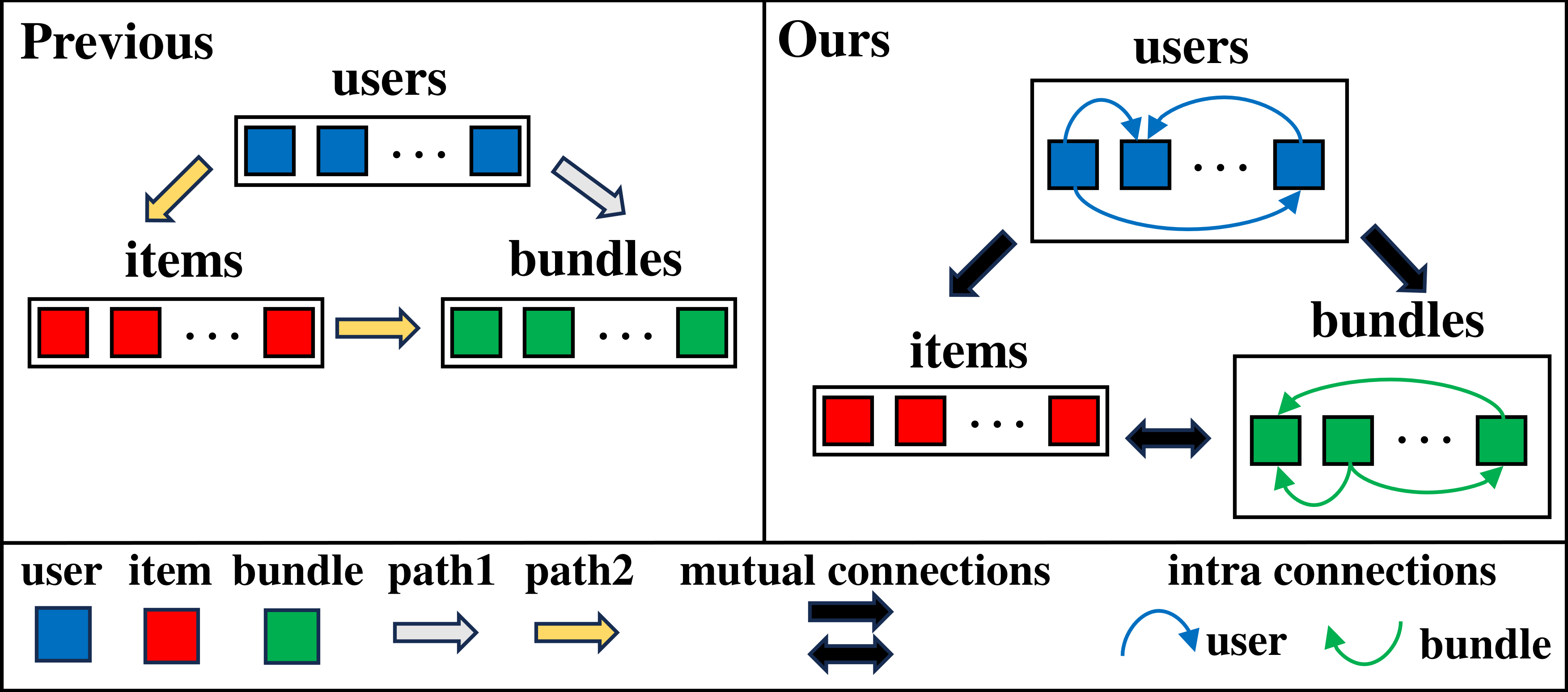} 
	\caption{Comparison of relation modeling. Blue, red, and green squares denote users, items, and bundles. Gray and gold arrows show paths used in prior methods; black bidirectional arrows show direct relations in HED. Blue and green curves denote intra-user and intra-bundle relations, respectively.}
	\label{fig:Bundle}
\end{figure}

Unlike single-item recommendation, bundle recommendation must represent a triadic relationship and use item-level evidence while ultimately ranking bundles. Existing methods address this challenge through separate user--bundle, user--item, and bundle--item views, relational graphs, or a unified hypergraph~\cite{chang2020bundle,tan2021intention,wang2021relational,avny2022bruce,ma2022crosscbr,zhao2022multi,yu2022unifying,wei2023strategy}. These designs are effective, but separating the views can lengthen the path through which item information reaches bundle scores, while a unified representation can underuse some available relations. Figure~\ref{fig:Bundle} summarizes this distinction. We therefore seek a representation that retains the cross-type relations among users, items, and bundles while modeling useful intra-user and intra-bundle structure.

To address this gap, HED comprises an embedding initialization layer, a dual convolutional layer, and a prediction layer. The first branch propagates information over the complete hypergraph, whereas the second propagates over the user--bundle graph to preserve ranking-specific signals. The prediction layer scores candidate bundles from the resulting user and bundle representations.
Our contributions are as follows:
\begin{itemize}
	\item We construct a complete hypergraph that retains observed cross-type interactions among users, items, and bundles, augmented with intra-user and intra-bundle relations.
	\item We propose a dual-propagation architecture that fuses complete-hypergraph context with a user--bundle branch, so item-aware information can improve the representations used for bundle ranking.
	\item We evaluate HED using Recall and NDCG at three cutoffs on two public datasets, supported by ablations, sensitivity analyses, and resource measurements. The results identify both consistent gains over the evaluated baselines and the model's memory limitations.
\end{itemize}

\section{Related Work}
\label{sec:RelatedWorks}

Bundle recommendation methods differ in how explicitly they model item-level evidence. Early approaches treat a bundle as an atomic object and learn preferences with factorization or sequential recommendation~\cite{rendle2010factorizing}. Later methods incorporate the items contained in a bundle: Cao et al.~jointly model preferences for items and user-generated lists~\cite{cao2017embedding}; Chen et al.~aggregate item embeddings with attention and train user--item and user--bundle objectives jointly~\cite{chen2019matching}; and Tan et al.~disentangle item-level from bundle-level preference~\cite{tan2021intention}. BRUCE further uses contextualized item embeddings and self-attention to model both item-specific and bundle-level preferences~\cite{avny2022bruce}. These methods establish the value of item information, but they do not uniformly retain all user--item--bundle relations in one propagation structure.

\begin{figure*}[!t]
\centering
\includegraphics[width=0.95\textwidth]{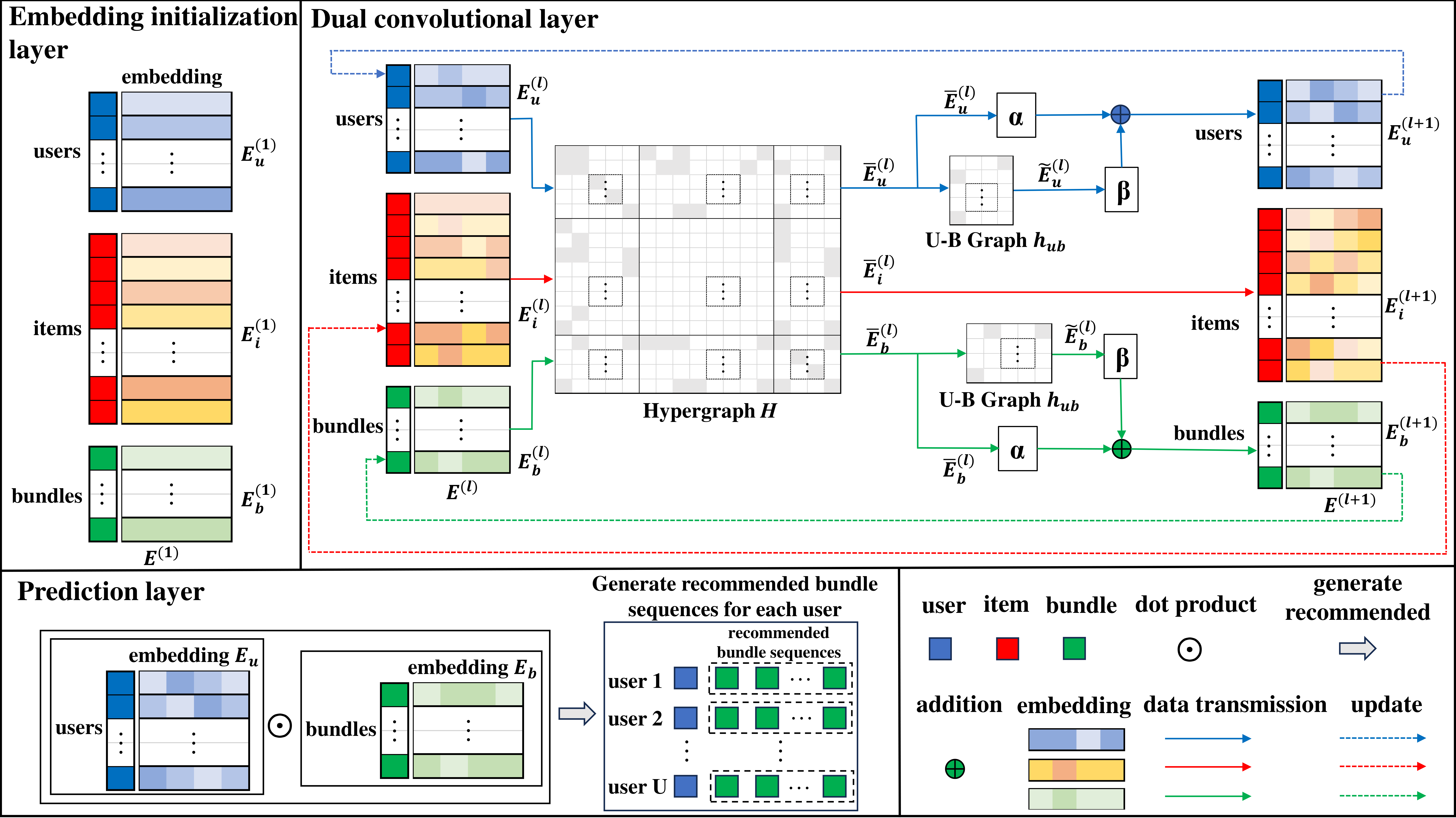} 
\caption{Overview of HED. The embedding initialization, dual convolutional, and prediction layers are described in Sections~\ref{em}, \ref{DC}, and~\ref{pr}, respectively. Blue, red, and green squares denote users, items, and bundles; colored rectangles are their embeddings. Solid and dashed lines show information transmission and updates. The hypergraph $H$ is defined in Section~\ref{HG}; dashed boxes in the prediction layer denote generated recommendation sequences.}
\label{fig:framework}
\end{figure*} 

Graph-based models preserve higher-order collaborative signals in different ways. Chang et al.~use heterogeneous graphs and graph convolution~\cite{chang2020bundle}; Wang et al.~augment user and bundle representations with neighbor interactions~\cite{wang2021relational}; Ma et al.~connect bundle and item views through contrastive learning~\cite{ma2022crosscbr}; and Zhao et al.~separate global and local user intent~\cite{zhao2022multi}. 

Yu et al.~unify multiple associations in a hypergraph~\cite{yu2022unifying}, whereas Wei et al.~model bundle strategies with a graph transformer~\cite{wei2023strategy}.
HED is closest to UHBR in its use of a hypergraph, but differs in three respects: it explicitly retains user--item and bundle--item blocks together with intra-user and intra-bundle relations; it propagates embeddings for items as well as users and bundles; and it preserves a dedicated user--bundle branch for the ranking objective. The experiments therefore test whether this combination improves bundle ranking, rather than claiming that graph- or transformer-based alternatives are universally inferior.

\section{Problem Formulation}
\label{sec:PF}

We study implicit-feedback bundle recommendation with bundles $S_B=\{b_i\mid i=1,\ldots,B\}$, users $S_U=\{u_j\mid j=1,\ldots,U\}$, and items $S_I=\{i_k\mid k=1,\ldots,I\}$. The observed relations are represented by three matrices: $A_{ub}$ records user--bundle interactions, $A_{ui}$ records user--item interactions, and $A_{bi}$ records bundle membership. The binary matrices have dimensions $A_{ub}\in\{0,1\}^{U\times B}$, $A_{ui}\in\{0,1\}^{U\times I}$, and $A_{bi}\in\{0,1\}^{B\times I}$.

The task is to rank bundles for each user from the observed user--bundle, user--item, and bundle--item relations. Given the three matrices as input, the model produces a score for a user--bundle pair:

\begin{figure}[!t]
	\centering
\includegraphics[width=0.98\linewidth]{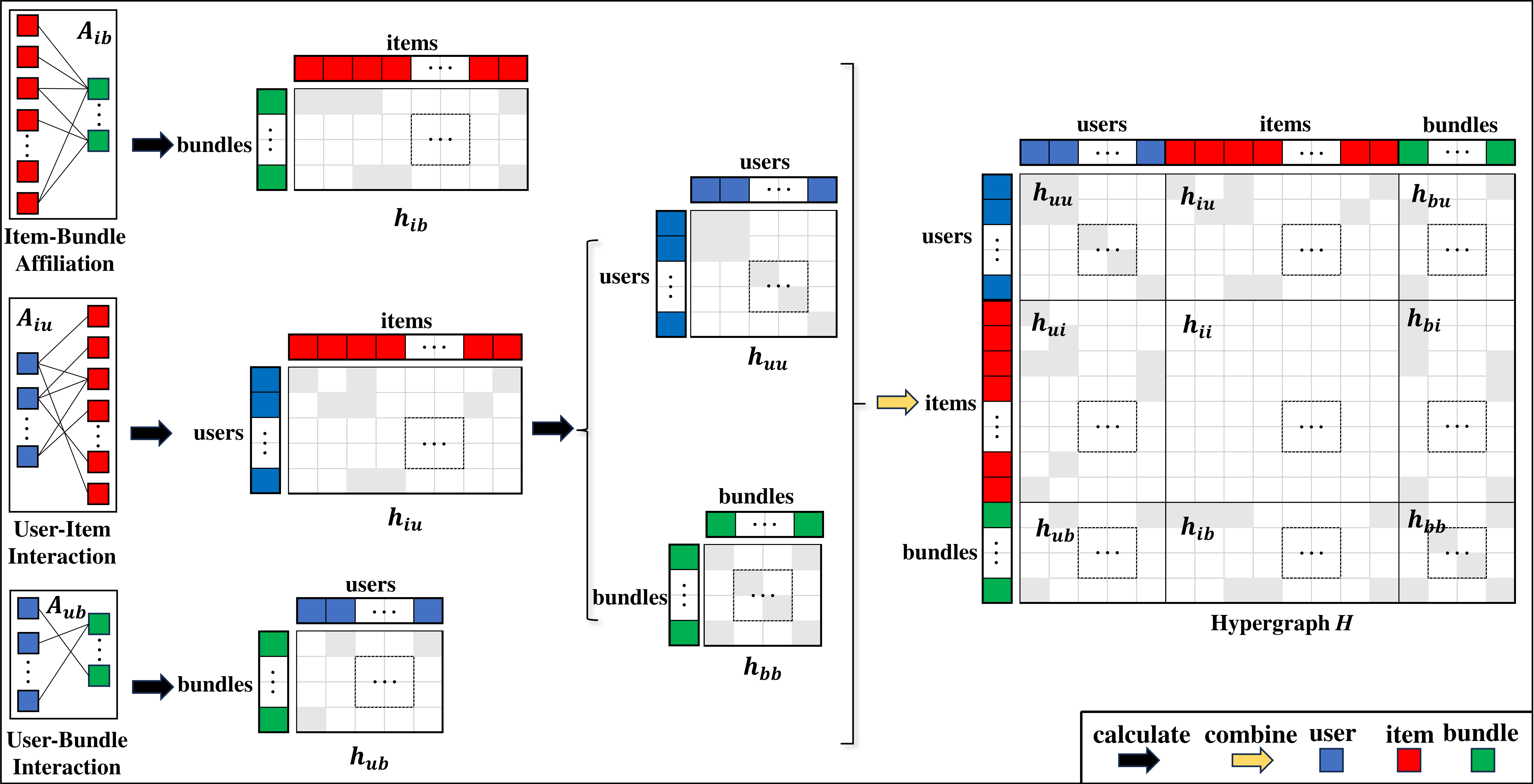} 
\caption{Construction of the complete hypergraph $H$. Blue, red, and green squares denote users, items, and bundles. From $A_{ub}$, $A_{ui}$, and $A_{bi}$, we construct the U-B, U-I, and B-I blocks; $h_{uu}$ and $h_{bb}$ are derived from $h_{ub}$, whereas $h_{ii}$ is the zero matrix. Black and gold arrows denote computation and synthesis. Ellipses indicate omitted entities and interactions.}
	\label{fig:HyperGraph}
\end{figure} 

\begin{equation}\label{eq1}
\hat{y} = f(u, b\mid A_{ub}, A_{ui}, A_{bi}; \theta).
\end{equation}
Here, $\theta$ denotes the trainable parameters and $\hat{y}$ is the score assigned to user $u$ and bundle $b$. Because the data provide implicit feedback, this score is used for ranking rather than interpreted as a calibrated purchase probability.

\section{Methodology}

HED addresses the ranking problem through the three-stage framework shown in Figure~\ref{fig:framework}: embedding initialization, dual convolution, and prediction. Section~\ref{HG} defines the complete hypergraph, and Section~\ref{NN} describes the propagation and scoring networks.

\subsection{Complete Hypergraph Structure}
\label{HG}

We use a neighborhood-incidence hypergraph over $N=U+I+B$ entities. Rows index vertices and columns index entity-centered hyperedges: a column contains the vertices related to its center through the retained blocks. A column with multiple nonzero entries therefore supports joint aggregation over a neighborhood. Here, ``complete'' denotes coverage of the specified relation types, not an all-to-all graph or the inclusion of every possible hyperedge.

UHBR~\cite{yu2022unifying} provides an important starting point by unifying user, item, and bundle associations in a hypergraph. Our construction extends this formulation in three ways: it includes the user--item and bundle--item blocks explicitly, adds intra-user and intra-bundle relations, and keeps an item embedding block for propagation. Figure~\ref{fig:HyperGraph} illustrates the construction. The goal is not to add relations indiscriminately, but to retain the interactions available in the data while keeping the final score focused on users and bundles.

\begin{equation}\label{eqH1}
	H = \begin{bmatrix}
		h_{uu} & h_{iu} & h_{bu}\\
		h_{ui} & h_{ii} & h_{bi}\\
		h_{ub} & h_{ib} & h_{bb}\\
	\end{bmatrix}
\end{equation}

To make the block orientation explicit, $h_{xy}$ has rows of type $y$ and columns of type $x$. Thus, $h_{ub}=A_{ub}^{\top}$, $h_{iu}=A_{ui}$, and $h_{ib}=A_{bi}$; the opposite-direction blocks are their transposes. The user and bundle shared-neighbor counts are $C_u=A_{ub}A_{ub}^{\top}$ and $C_b=A_{ub}^{\top}A_{ub}$. 

Off-diagonal entries of $h_{uu}$ and $h_{bb}$ are one when the corresponding count exceeds $n$, and zero otherwise; their diagonals are one. The dataset-specific $n$ is reported in Table~\ref{parameters}.
Shared-neighbor thresholding retains repeated collaborative evidence while suppressing isolated overlaps. It introduces no extra learnable relation parameters and controls the number of retained intra-type entries. Its simplicity also has a cost: raw counts favor active users and popular bundles, so one threshold need not transfer across datasets. Normalized overlap can reduce degree bias, weighted similarities can retain graded evidence, content-based semantic similarity between item or bundle descriptions could supply relations that interaction counts miss~\cite{ngulea,dhegra,chalea}, and learnable strengths can adapt to the ranking objective, but each changes the construction or optimization. These alternatives are not compared here and remain future work.

The main configuration uses $h_{ii}=0$. This excludes direct item-centered self-incidence in that block; it does not remove item information from the cross-type blocks or prevent indirect item--item communication through shared hyperedges. Section~\ref{II} compares this choice with $h_{ii}=\mathrm{Id}_I$.\subsection{Network Structure}
\label{NN}
\subsubsection{Embedding initialization layer}
\label{em}
We consider implicit feedback, so the model receives interaction histories rather than explicit ratings or side features. We initialize trainable embeddings for users, items, and bundles and stack them in the same order as the node blocks in the complete hypergraph. For embedding dimension $d$, the three blocks are $E_u^{(1)} \in \mathbb{R}^{U \times d}$, $E_i^{(1)} \in \mathbb{R}^{I \times d}$, and $E_b^{(1)} \in \mathbb{R}^{B \times d}$, yielding $E^{(1)} \in \mathbb{R}^{(U+I+B) \times d}$:

\begin{equation}\label{eq2}
	E^{(1)}= \begin{bmatrix}
		E_u^{(1)} \\
		E_i^{(1)} \\
		E_b^{(1)}
	\end{bmatrix}
\end{equation}

\subsubsection{Dual convolutional layer}
\label{DC}
Graph and hypergraph convolution update node representations by propagating messages through the corresponding relation structure~\cite{wang2020disentangled,feng2019hypergraph}. Following the lightweight propagation principle used in recommender systems~\cite{he2020lightgcn} and the efficient hypergraph convolution of UHBR~\cite{yu2022unifying}, we use normalized propagation without feature transformations or nonlinear activations:
\begin{equation}\label{eq3}
	\begin{split}
		& \hat{H} = D_V^{-1/2}HD_E^{-1/2}H^\text{T}D_V^{-1/2} \\
		& E^{(l+1)} = \hat{H}E^{(l)}     
	\end{split}
\end{equation}
where $D_V$ and $D_E$ are the node- and hyperedge-degree matrices of $H$, respectively, and $\hat{H}$ is the normalized propagation matrix. $E^{(l)}$ and $E^{(l+1)}$ denote the embeddings at layers $l$ and $l+1$.

The complete-hypergraph branch first produces $\bar E^{(l)}=\hat H E^{(l)}$, with user, item, and bundle blocks. The U-B branch then refines the user and bundle blocks using only their observed interactions. Define the same normalized incidence operator as in Equation~(\ref{eq3}):
\begin{equation}\label{eq:operator}
\mathcal{P}(X)=D_r(X)^{-1/2}X D_c(X)^{-1/2}X^{\top}D_r(X)^{-1/2},
\end{equation}
where $D_r(X)$ and $D_c(X)$ contain row and column sums, respectively; inverse degrees are zero for isolated entities. With the two U-B orientations, $P_u=\mathcal P(A_{ub})\in\mathbb R^{U\times U}$ and $P_b=\mathcal P(A_{ub}^{\top})\in\mathbb R^{B\times B}$. Writing the user and bundle operators separately avoids applying one matrix to blocks of different sizes:
\begin{equation}\label{eq4}
\begin{aligned}
\bar E^{(l)} &= \hat H E^{(l)},\\
\tilde E_u^{(l)} &= P_u\bar E_u^{(l)},\quad
\tilde E_b^{(l)} = P_b\bar E_b^{(l)},\\
E_u^{(l+1)} &= \alpha\bar E_u^{(l)}+\beta\tilde E_u^{(l)},\\
E_i^{(l+1)} &= \bar E_i^{(l)},\\
E_b^{(l+1)} &= \alpha\bar E_b^{(l)}+\beta\tilde E_b^{(l)}.
\end{aligned}
\end{equation}
The two pathways are coupled rather than independent views: the U-B branch operates on representations already enriched by items. Its role is to refine ranking-related neighborhoods, while the item block bypasses this refinement. The fixed weighted addition preserves dimension $d$ and adds no feature-transformation parameters; aggregating complementary evidence sources in this way has also proved effective in other prediction tasks~\cite{maamul}. Unlike concatenation, it does not enlarge the score representation. HED's contribution is this relation construction and task-specific coupling of established propagation operators, rather than a new convolution primitive.

Because different propagation depths capture different neighborhoods, using only the final layer can discard useful intermediate signals. Following LightGCN~\cite{he2020lightgcn}, we aggregate the layer-wise user and bundle embeddings with depth-dependent weights without increasing their dimensionality:

\begin{equation}\label{eq5}
\begin{split}
E_u = \sum_{l=1}^L \frac{1}{l+1}E_u^{(l)} \\ E_b = \sum_{l=1}^L \frac{1}{l+1}E_b^{(l)} 
\end{split}
\end{equation}
where $L$ is the total number of layers.

\subsubsection{Prediction layer}
\label{pr}
After obtaining the final user and bundle representations, the preference score for user $u$ and bundle $b$ is computed as follows:
\begin{equation}\label{eq6}
	\hat{y}_{ub} = e_u^\top e_b
\end{equation}
where $\top$ denotes transposition. For each user, candidate bundles are ranked by $\hat{y}_{ub}$ to form the recommendation list.

\subsubsection{Model Training}
We optimize the model with the User Interest Boundary (UIB) loss~\cite{zhuo2022learning}. UIB introduces a user-specific boundary that separates observed from unobserved interactions:
\begin{equation}\label{eq7}
	b_u = W^{\top}e_u
\end{equation}
where $e_u\in\mathbb{R}^d$ is the representation of user $u$ and $W\in\mathbb{R}^d$ is learnable. The boundary loss penalizes observed scores below the boundary and sampled unobserved scores above it:
\begin{equation}\label{eq8}
	\mathcal{L} = \sum_{(u,p)\in R^+}\phi(b_u-\hat y_{up})\; +\!\sum_{(u,n)\in R^-}\phi(\hat y_{un}-b_u)  
\end{equation}
where $\phi$ is the increasing surrogate penalty used by UIB, $R^+$ and $R^-$ denote the observed and sampled unobserved interaction sets, respectively, and $\hat{y}_{ub}^{p}$ and $\hat{y}_{ub}^{n}$ are their predicted scores. We minimize this loss with AdamW~\cite{loshchilov2017decoupled}.

\begin{table}[!t]
\scriptsize
	\centering
	\caption{Dataset statistics.}
	\label{tab1}
	\begin{tabular}{ccc}
		\hline
		Dataset             & NetEase & Youshu \\
		\hline
		\#U               & 18528   & 8039   \\
		\#I                & 123628  & 32770  \\
		\#B              & 22864   & 4771   \\
		\hline
		\#U-B         & 302303  & 51337  \\
		\#U-I           & 1128065 & 138515 \\
		\#B-I         & 1778838 & 176667 \\
		\hline
		U-B Density & 0.07\%  & 0.13\% \\
		U-I Density   & 0.05\%  & 0.05\% \\
		B-I Density & 0.06\%  & 0.11\% \\
		\hline
	\end{tabular}
\end{table}

\section{Experiments}
\label{sec:experiments}
We evaluate HED from three complementary perspectives: predictive performance, component contribution, and computational cost.

\subsection{Experimental Setup}
\noindent\textbf{Datasets.} We use two public bundle-recommendation datasets from different domains: NetEase\footnote{https://music.163.com/}, which contains user, music, and music-bundle interactions, and Youshu\footnote{https://www.yousuu.com/}, which contains user, book, and book-bundle interactions. Their sizes and interaction densities are summarized in Table~\ref{tab1}. We also attempted iFashion, but the current implementation exceeded the available GPU memory; consequently, no ranking result is reported for that dataset.

\noindent\textbf{Evaluation Metrics.} We report Recall@k and NDCG@k, which measure retrieval coverage and ranking quality, respectively. We evaluate each metric at $k\in\{20,40,80\}$ to examine both short-list and longer-list behavior.

\noindent\textbf{Baselines.} The comparison covers six representative methods: DAM\cite{chen2019matching}, which uses attention and multi-task learning; RGNN\cite{wang2021relational}, which models relational neighbor interactions; MIDGN\cite{zhao2022multi}, which disentangles global and local intent; CrossCBR\cite{ma2022crosscbr}, which performs cross-view contrastive learning; UHBR\cite{yu2022unifying}, which uses a unified hypergraph; and BundleGT\cite{wei2023strategy}, which uses a graph transformer. This set spans attention, relational graph, contrastive, hypergraph, and transformer-based approaches.

\noindent\textbf{Training Details.} We use the reported 80/20 train--validation split for HED. The metrics reflect this validation protocol; a separate untouched test partition and repeated-seed uncertainty are not included. Experiments run on an Intel Xeon Platinum 8336C CPU and an NVIDIA GeForce RTX 4090 GPU. We select the learning rate from $\{5\times10^{-1},5\times10^{-2},5\times10^{-3},5\times10^{-4},5\times10^{-5}\}$ and use $5\times10^{-3}$ in the reported runs; the remaining dataset-specific settings are listed in Table~\ref{parameters}.

\subsection{Comparative Results}
We reproduce the baselines using publicly available implementations when possible and otherwise follow the corresponding published descriptions. HED is evaluated with embedding sizes 32, 64, and 128; the default size is dataset-specific as shown in Table~\ref{parameters}. Because resource requirements differ across methods, the predictive results should be read together with Table~\ref{costs}.

\begin{table}[!t]
\scriptsize
    \centering
    \caption{Hyperparameter settings.}
    \label{parameters}
\setlength{\tabcolsep}{2pt}
\begin{tabular}{ccc}
\hline
Parameter & NetEase      & Youshu     \\ \hline
Embedding Size                 & 128          & 64         \\
$n$             & 3            & 10         \\
Epochs                         & 500          & 300        \\
Batch Size                     & 2048         & 1024       \\
Hypergraph Convolution Dropout & 0.1          & 0.2        \\ \hline
$L$              & \multicolumn{2}{c}{2}     \\
$\alpha$          & \multicolumn{2}{c}{1/2}   \\
$\beta$           & \multicolumn{2}{c}{1/100} \\
Learning Rate                  & \multicolumn{2}{c}{5e-3}  \\
U-B graph Convolution Dropout  & \multicolumn{2}{c}{0.01}  \\
L2-norm                        & \multicolumn{2}{c}{0.1}  \\ \hline
\end{tabular}
\end{table}

Tables~\ref{tab2.1} and~\ref{tab2.2} compare HED with the baselines. Underlined entries mark the best baseline result, bold entries mark the best HED variant, and the final row reports the improvement over the best baseline. The number after HED denotes the embedding dimension. The setup follows Table~\ref{parameters}, except that HED-32 is trained for 500 epochs on Youshu.

\begin{table}[!t]
\scriptsize
 \centering
\caption{Performance comparison on NetEase.}
	\label{tab2.1}
\setlength{\tabcolsep}{1.2pt}
\begin{tabular}{ccccccc}
\hline
Model     & Recall@20 & NDCG@20 & Recall@40 & NDCG@40 & Recall@80 & NDCG@80 \\ \hline
DAM       & 0.0411    & 0.021   & 0.069     & 0.0281  & 0.109     & 0.0372  \\
RGNN      & 0.0644    & 0.0327  & 0.1023    & 0.0431  & 0.1602    & 0.0551  \\
MIDGN     & 0.0678    & 0.0343  & 0.1085    & 0.0451  & 0.1654    & 0.0578  \\
CrossCBR  & 0.0842    & 0.0457  & 0.1264    & 0.0569  & 0.1805    & 0.0679  \\
UHBR      & 0.0853    & 0.0452  & 0.1335    & 0.0577  & \underline{0.1992} & \underline{0.0732}  \\
BundleGT  & \underline{0.0903} & \underline{0.0478} & \underline{0.1389} & \underline{0.0606} & 0.1655 & 0.0572 \\ \hline
HED-32    & 0.0895    & 0.0478  & 0.1402    & 0.0608  & 0.2099    & 0.0764  \\
HED-64    & 0.0934    & 0.0496  & 0.1428    & 0.0624  & 0.2123    & 0.0776  \\
\textbf{HED-128}   & \textbf{0.0963} & \textbf{0.051}  & \textbf{0.1459} & \textbf{0.0638} & \textbf{0.213}  & \textbf{0.0783} \\ \hline
\textbf{\%Improv.} & \textbf{6.64\%}    & \textbf{6.69\%}  & \textbf{5.04\% }   & \textbf{5.28\%}  & \textbf{6.93\%}    & \textbf{6.97\%}  \\ \hline
\end{tabular}
\end{table}

\begin{table}[!t]
\scriptsize
 \centering
\caption{Performance comparison on Youshu.}
	\label{tab2.2}
\setlength{\tabcolsep}{1.2pt}
\begin{tabular}{ccccccc}
\hline
Model                         & Recall@20 & NDCG@20 & Recall@40 & NDCG@40 & Recall@80 & NDCG@80 \\ \hline
DAM                           & 0.2082    & 0.1198  & 0.289     & 0.1418  & 0.3915    & 0.1658  \\
RGNN                          & 0.2732    & 0.1571  & 0.3814    & 0.1857  & 0.4937    & 0.2116  \\
MIDGN                         & 0.2682    & 0.1527  & 0.3712    & 0.1808  & 0.4817    & 0.2063  \\
CrossCBR                      & 0.2813    & 0.1668  & 0.3785    & 0.1938  & 0.4943    & 0.2198  \\
UHBR                          & \underline{0.3049} & \underline{0.1821} & \underline{0.4166} & \underline{0.2122} & \underline{0.5308} & \underline{0.2383}  \\
BundleGT                      & 0.2905    & 0.1737  & 0.3909    & 0.2011  & 0.5124    & 0.2242  \\ \hline
HED-32                        & 0.3157    & 0.1886  & 0.4223    & 0.2175  & 0.5412    & 0.2445  \\
\textbf{HED-64}    & \textbf{0.3165} & \textbf{0.1904} & \textbf{0.4244} & \textbf{0.2197} & \textbf{0.5427} & \textbf{0.2466} \\
HED-128                       & 0.3122    & 0.1885  & 0.421     & 0.2176  & 0.5373    & 0.2438  \\ \hline
\textbf{\%Improv.} & \textbf{3.80\% }   & \textbf{4.56\%}  & \textbf{1.87\%}    & \textbf{3.53\%}  & \textbf{2.24\%}    & \textbf{3.48\%} \\ \hline
\end{tabular}
\end{table}

\noindent\textbf{Overall comparison.} HED-128 is the strongest HED variant on NetEase, whereas HED-64 is strongest on Youshu. Relative to the strongest baseline in each metric, the improvement ranges from 5.04\% to 6.97\% on NetEase and from 1.87\% to 4.56\% on Youshu. Within these two datasets, the results support the proposed representation and show that the preferred embedding size is dataset-dependent.

\noindent\textbf{Effect of dataset structure.} The larger gains on NetEase coincide with a larger and sparser interaction structure, as summarized in Table~\ref{tab1}. This pattern is consistent with the hypothesis that the complete hypergraph can provide useful auxiliary paths when direct user--bundle observations are sparse. Since only two datasets are evaluated, this should be treated as an observed pattern rather than evidence of universal scalability.

\noindent\textbf{Ranking quality.} The relative improvements in NDCG exceed those in Recall at all three cutoffs on both datasets; the gap ranges from 0.04 to 1.66 percentage points. This pattern is consistent with improved ordering of retrieved bundles in addition to greater coverage. A plausible explanation is that the complete-hypergraph branch supplies item-aware context, while the U-B branch keeps the propagated representations aligned with the final bundle-ranking objective.

\subsection{Ablation Study}

We use four ablations to isolate the U-B branch and intra-type blocks: HED$-$c removes U-B graph convolution; HED$-$cu additionally sets $h_{uu}=0$; HED$-$cb additionally sets $h_{bb}=0$; and HED$-$cbu sets both $h_{uu}=0$ and $h_{bb}=0$.

\begin{table*}[!t]
\centering
\scriptsize
\caption{Ablation results on Youshu.}
\label{tab3}
\setlength{\tabcolsep}{4pt}
\begin{tabular}{cccccc}
\hline
          & HED            & HED-c            & HED-cu           & HED-cb           & HED-cbu          \\ \hline
Recall@20 & 0.3165 (100\%) & 0.3086 (96.94\%) & 0.2987 (94.38\%) & 0.2965 (93.68\%) & 0.3011 (95.13\%) \\
NDCG@20   & 0.1904 (100\%) & 0.1841 (96.69\%) & 0.1776 (93.28\%) & 0.1779 (93.43\%) & 0.1792 (94.12\%) \\
Recall@40 & 0.4244 (100\%) & 0.4164 (98.11\%) & 0.4044 (95.29\%) & 0.4025 (94.84\%) & 0.4046 (95.33\%) \\
NDCG@40   & 0.2197(100\%)  & 0.2128 (96.86\%) & 0.2057 (93.63\%) & 0.2064 (93.95\%) & 0.2076 (94.49\%) \\
Recall@80 & 0.5427 (100\%) & 0.5304 (97.73\%) & 0.5235 (96.46\%) & 0.5267 (97.05\%) & 0.5254 (96.81\%) \\
NDCG@80   & 0.2466 (100\%) & 0.2397 (97.20\%) & 0.2323 (94.20\%) & 0.2337 (94.77\%) & 0.2347 (95.17\%) \\ \hline
\end{tabular}
\end{table*}

HED$-$c is the complete-hypergraph-only variant, so the existing ablation already tests that pathway without U-B refinement. Table~\ref{tab3} reports each absolute metric followed by its percentage relative to the full HED model. The six metrics of HED$-$c are all lower than those of HED, showing that the U-B branch contributes complementary ranking information. Removing either $h_{uu}$ or $h_{bb}$ causes further degradation in most metrics. Setting both blocks to zero does not recover the full model and is not uniformly better than removing only one block, indicating that the two intra-type relations have complementary rather than simply additive effects. These removals do not change the embedding dimension or trainable entity tables; they change relation coverage and propagation. However, there is no U-B-only result, concatenation comparison, or fully matched-budget fusion study. The ablations support the retained components but do not establish optimality over these alternatives.

\begin{figure}[!t]
	\centering
	\includegraphics[width=\linewidth]{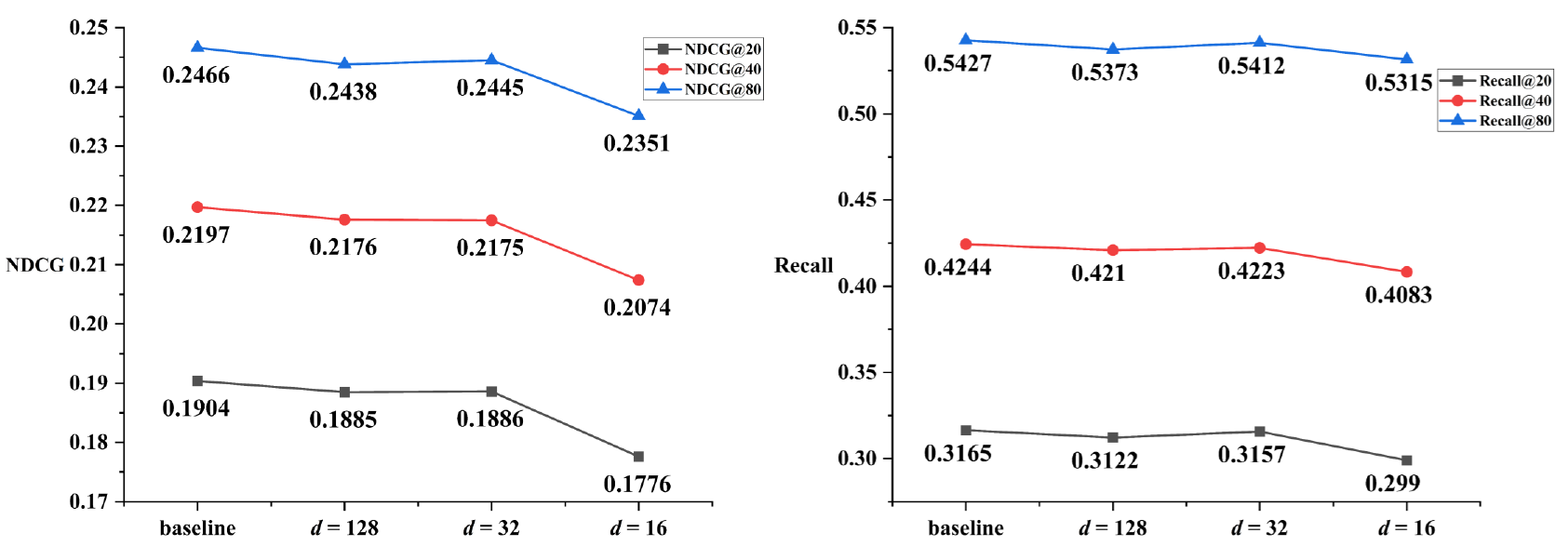} 
	\caption{Recall and NDCG performance under different $d$}
	\label{fig:ES}
\end{figure} 

\begin{figure}[!t]
	\centering
	\includegraphics[width=0.98\linewidth]{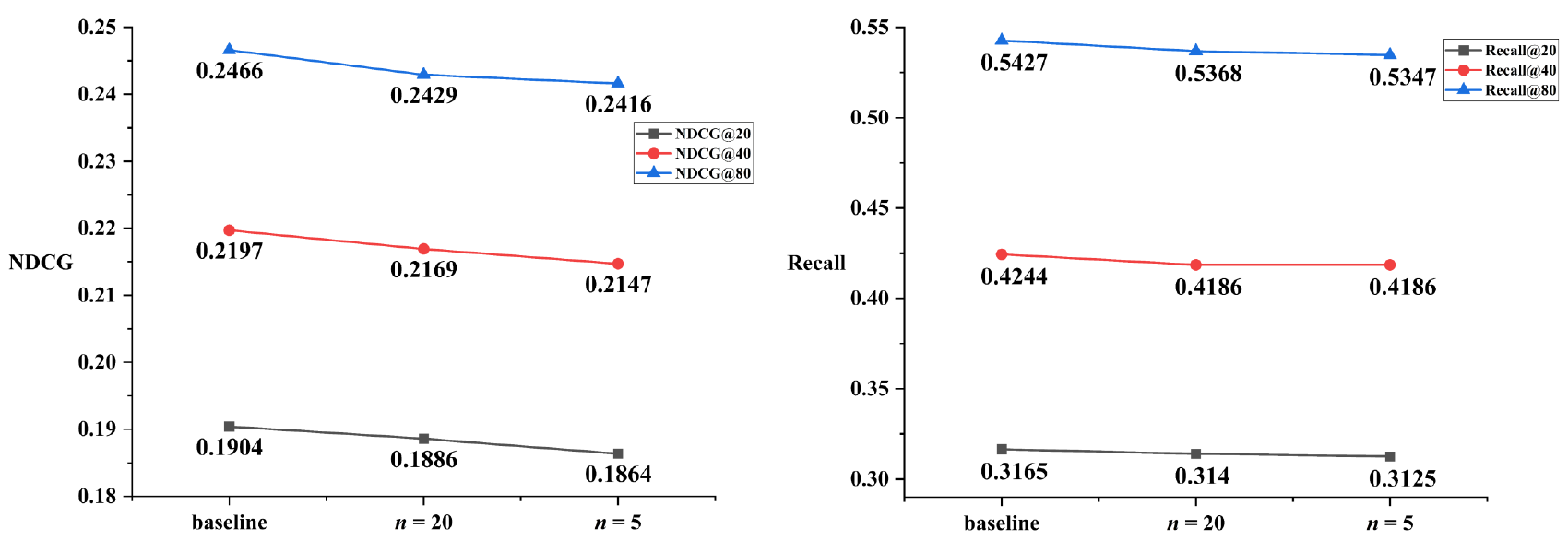} 
	\caption{Recall and NDCG performance under different $n$}
	\label{fig:n}
\end{figure} 

\begin{figure}[!t]
	\centering
	\includegraphics[width=0.98\linewidth]{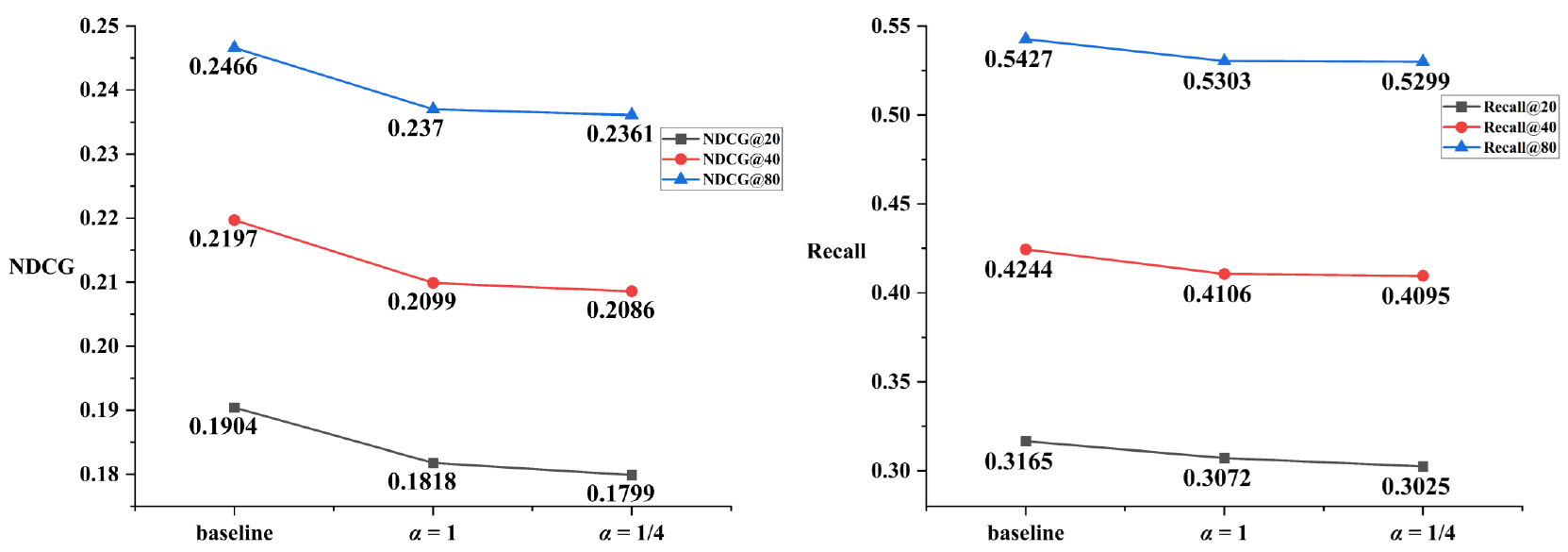} 
	\caption{Recall and NDCG performance under different $\alpha$.}
	\label{fig:alpha}
\end{figure} 

\subsection{Sensitivity Analysis}

\begin{figure}[!t]
\centering
	\includegraphics[width=\linewidth]{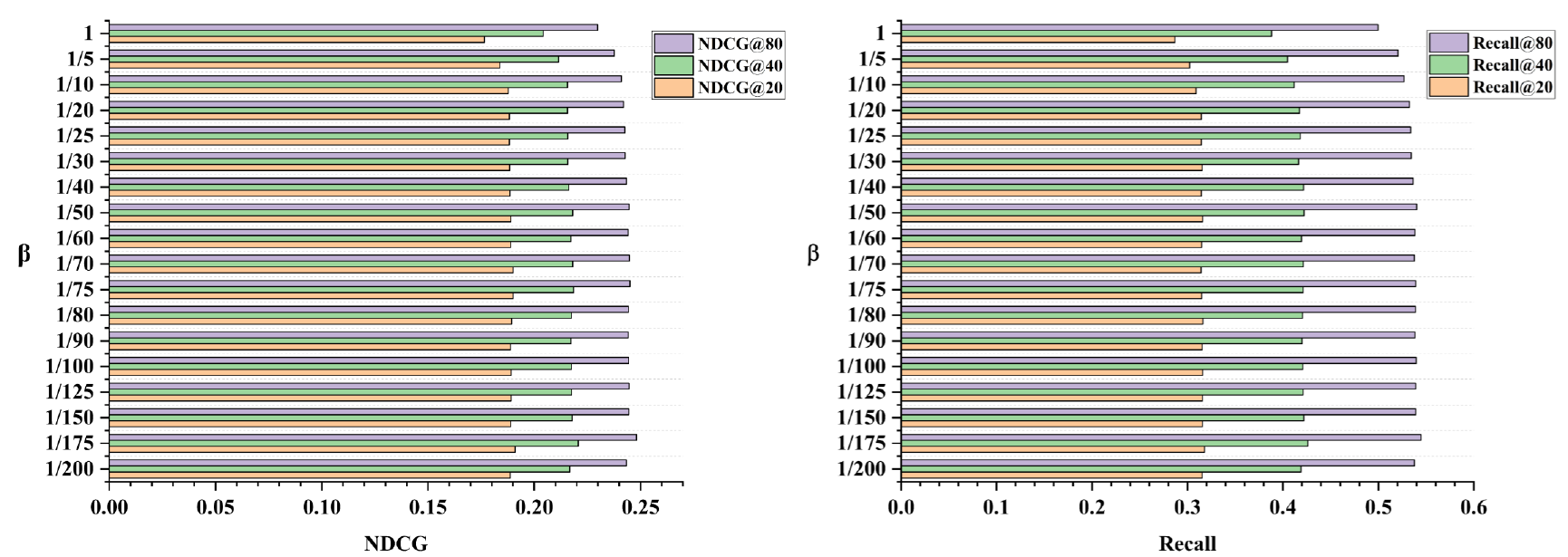} 
	\caption{Recall and NDCG performance under different $\beta$}
	\label{fig:beta}
\end{figure} 
We next examine the sensitivity of HED to the embedding dimension $d$, the intra-type threshold $n$, and the fusion weights $\alpha$ and $\beta$. Each sweep varies one parameter while fixing the remaining settings, as summarized in Figures~\ref{fig:ES}--\ref{fig:beta}.

\noindent\textbf{Embedding dimension.} $d$ is the embedding dimension for all three entity types. With the baseline $d=64$, $d=16$ produces a clear performance drop, whereas $d\geq32$ remains close to the baseline and improves some individual metrics. This suggests that very small embeddings cannot retain all signals propagated from the complete hypergraph and the U-B graph, while increasing $d$ beyond 32 yields diminishing returns in the reported sweep.
    
\noindent\textbf{Intra-type threshold.} $n$ controls how many shared neighbors are required to create an intra-user or intra-bundle edge. The baseline uses $n=10$ on Youshu and $n=3$ on NetEase. Halving or doubling $n$ causes only modest changes in the reported metrics; the largest observed decline is 1.48\% in NDCG@20 at $n=5$. Thus, the model is not highly sensitive in the tested range, but the appropriate threshold remains dataset-dependent.

\noindent\textbf{Fusion weights.} $\alpha$ weights the complete-hypergraph branch, with the baseline set to $\alpha=1/2$. The sweep shows lower performance at both $\alpha=1/4$ and $\alpha=1$, showing that rescaling the complete-hypergraph contribution affects performance. This sweep alone does not establish an equal contribution from the two branches.
$\beta$ weights the U-B branch and is set to $1/100$ in the baseline. The broader sweep in Figure~\ref{fig:beta} shows a clear regime change: values $\beta\geq1/20$ degrade performance, whereas smaller values are comparatively stable. Together with HED$-$c, this supports interpreting the U-B pathway as a useful correction to a dominant complete-hypergraph encoder. The numerical ratio $\beta/\alpha=0.02$ is not a percentage attribution: the output also depends on operator normalization, embedding norms, and repeated propagation.

\subsection{Effect of the item--item block}
\label{II}
We compare the zero item--item block with an identity block on Youshu (Table~\ref{tab:ii}). Identity incidence adds item self-information; it is not an item-similarity graph and creates no direct off-diagonal item--item relations. The identity variant improves Recall@80 (0.5392 versus 0.5377), whereas the zero-block reference is higher on the other five metrics. These differences are small and have no reported uncertainty, so they provide empirical motivation for the zero block rather than a theoretical guarantee. The reference scores in this comparison differ slightly from the main HED-64 results; without matched-seed uncertainty, these small differences should be interpreted cautiously.

\begin{table}[!t]
\scriptsize
\caption{Item-block comparison on Youshu (separately reported reference).}
\label{tab:ii}
	\centering
\begin{tabular}{ccc}
\hline
& $h_{ii}=\mathrm{Id}_I$ & $h_{ii}=0$ reference   \\
\hline
Recall@20 & 0.3152 & 0.3153 \\
NDCG@20   & 0.1896 & 0.1903 \\
Recall@40 & 0.4208 & 0.423  \\
NDCG@40   & 0.2184 & 0.2197 \\
Recall@80 & 0.5392 & 0.5377 \\
NDCG@80   & 0.2446 & 0.2453 \\
\hline
\end{tabular}
\end{table}

\subsection{Implementation efficiency}
The complete hypergraph improves predictive performance at a nontrivial computational cost. Table~\ref{np} shows that HED-64 has a parameter count comparable to CrossCBR, exceeds those of UHBR and MIDGN, and remains below DAM and BundleGT on both datasets. Increasing the embedding dimension substantially increases the parameter count because HED maintains embeddings for users, items, and bundles; this explains the larger resource footprint of HED-128.

\begin{table}[!t]
\scriptsize
\caption{Parameter counts of compared methods.}
\label{np}
	\centering
\begin{tabular}{ccc}
\hline
Methods  & \#Params-NetEase & \#Params-Youshu \\ \hline
DAM      & 18490372         & 5047424         \\ 
BGCN     & 10586112         & 2941952          \\ 
RGNN     & 10561538         & 2917378         \\ 
MIDGN    & 4643776          & 836480          \\ 
CrossCBR & 10561280         & 2917120         \\ 
UHBR     & 2649152          & 819904          \\ 
BundleGT & 31683840         & 8751360         \\ 
HED-32   & 5280640          & 1458560         \\ 
HED-64   & 10561280         & 2917120         \\ 
HED-128  & 21122560         & 5834240         \\ \hline
\end{tabular}
\end{table}
Table~\ref{costs} compares the reported GPU memory. On NetEase, HED uses 23,148 MiB, about 2.75 times UHBR and 8.49 times CrossCBR, and also exceeds BundleGT's 10,432 MiB. On Youshu, HED uses 3,232 MiB, exceeding UHBR and CrossCBR but remaining below BundleGT. The memory reported for HED is unchanged across $d\in\{32,64,128\}$ at the available measurement precision, even though parameter counts grow. This is consistent with a substantial structure-related cost, but the measurements alone do not isolate its cause. MIDGN exceeds memory on NetEase in this resource comparison; its predictive scores elsewhere do not demonstrate feasibility in that configuration. The original timing records lack a recoverable unit and measurement definition, so we do not infer latency or throughput from them.

\begin{table}[!t]
\centering\scriptsize
\caption{Reported GPU memory (MiB). OOM: out of memory.}
\label{costs}
\begin{tabular}{lrr}\hline
Model & NetEase & Youshu\\\hline
MIDGN & OOM & 5,028\\
CrossCBR & 2,728 & 1,642\\
UHBR & 8,412 & 2,388\\
BundleGT & 10,432 & 5,392\\
HED-32 & 23,148 & 3,232\\
HED-64 & 23,148 & 3,232\\
HED-128 & 23,148 & 3,232\\\hline
\end{tabular}
\end{table}

\subsection{Complexity and scalability}
Let $m_{ub}$, $m_{ui}$, and $m_{bi}$ count observed interactions, $r_u$ and $r_b$ count retained entries of $h_{uu}$ and $h_{bb}$, and $s=\operatorname{nnz}(H)$. With symmetric cross-type blocks,
\begin{equation}\label{eq:nnz}
s=2(m_{ub}+m_{ui}+m_{bi})+r_u+r_b.
\end{equation}
Sparse storage of $H$ costs $O(s)$, and entity embeddings require $O(Nd)$ parameters. Materializing the normalized propagation matrix $\hat H$ can create many more entries than $H$; write $q=\operatorname{nnz}(\hat H)$. Applying this materialized operator costs $O(qd)$ per layer and requires $O(q)$ structure storage, with $q$ as large as $N^2$. A dense $N\times N$ float32 matrix alone would occupy approximately 101.4 GiB on NetEase ($N=165{,}020$) and 7.74 GiB on Youshu ($N=45{,}580$), before embeddings and training states. These are analytical dense-storage bounds, not measured allocations.

A factorized sparse application of Equation~(\ref{eq3}), multiplying successively by the degree factors, $H^{\top}$, and $H$, avoids explicitly forming $\hat H$. Its propagation cost is $O(sd)$ per layer. 

The two factorized U-B operators add $O(m_{ub}d)$, giving $O(L(s+m_{ub})d)$ for $L$ propagation steps and $O(s+m_{ub}+LNd)$ storage including layer activations; optimizer states add $O(Nd)$. This bound describes a possible sparse implementation, not a demonstrated speedup of the reported implementation. Scoring $K$ candidate bundles for one user adds $O(Kd)$, whereas scoring the entire catalog costs $O(Bd)$.

Construction can also be expensive. Enumerating shared-user pairs for each bundle and shared-bundle pairs for each user costs up to $O(\sum_b k_b^2+\sum_u k_u^2)$ candidate pair updates, where $k_b$ and $k_u$ are U-B degrees. Thresholding controls retained entries but does not automatically eliminate this candidate-generation cost. Intra-type blocks can approach $U^2+B^2$ entries in the worst case.

Potential remedies include factorized sparse propagation, degree-normalized overlap with capped neighborhoods, blockwise shared-neighbor construction, and sampled neighborhoods. Adaptive or learnable relation strengths could replace fixed thresholds. These changes require fresh accuracy--memory comparisons: none is claimed to resolve the iFashion failure in the present results.

\subsection{Scope and limitations}
The music and book datasets support HED in two observed domains, but do not establish generality for fashion or production-scale recommendation. We do not report an additional successful benchmark, repeated-seed confidence intervals, a U-B-only model, or a matched-budget concatenation study. At $d=64$, HED and CrossCBR have the same reported parameter count (Table~\ref{np}); this helps contextualize capacity but is not a controlled fusion ablation. Future evaluation should isolate relation coverage from representation capacity and report standardized resource measurements. No re-ranking deployment is evaluated here. HED also uses only implicit feedback; textual side information such as user reviews, analyzed with concept-level or document-level sentiment analysis~\cite{camsta,camkno,xiapol}, is a natural extension that we leave to future work.

\section{Conclusions}
We presented HED, a bundle-recommendation model that represents user--item--bundle interactions with a complete hypergraph and combines its propagation with a user--bundle convolution branch. The model initializes embeddings for all three entity types, uses the item-aware hypergraph to update user and bundle representations, and ranks bundles from the final user--bundle scores.

On NetEase, HED-128 improves over the strongest baseline by 5.04--6.97\% across the six reported metrics; on Youshu, HED-64 improves by 1.87--4.56\%. The ablation results support the contribution of the U-B branch and the intra-type relations, while the sensitivity analysis shows that moderate embedding dimensions and a dominant hypergraph branch with a small U-B correction is useful in the tested settings. The item-block comparison provides limited empirical support for omitting identity item incidence.

The results also make the limitations explicit. HED has a larger memory footprint than the most lightweight baselines, uses 23,148 MiB on NetEase in the reported runs, and runs out of memory on iFashion in the available environment. The evaluation is limited to two datasets, so the observed gains should not be interpreted as universal superiority or as proof of deployment scalability. Future work should reduce the cost of hypergraph construction and propagation, and should test the representation on broader bundle-recommendation settings.

\IEEEtriggeratref{15}
\bibliographystyle{IEEEtran}
\bibliography{references}
\end{document}